\RequirePackage{amsmath}
\documentclass{article} 
\usepackage{iclr2024_conference,times}

\usepackage{amsmath,amsfonts,bm}

\def\eqref#1{equation~\ref{#1}}

\def\1{\bm{1}}

\def\rr{{\textnormal{r}}}

\DeclareMathAlphabet{\mathsfit}{\encodingdefault}{\sfdefault}{m}{sl}
\SetMathAlphabet{\mathsfit}{bold}{\encodingdefault}{\sfdefault}{bx}{n}

\DeclareMathOperator*{\argmin}{arg\,min}

\usepackage{hyperref}
\usepackage{url}
\usepackage{booktabs}       
\usepackage{amsfonts}       
\usepackage{nicefrac}       
\usepackage{microtype}      
\usepackage{xcolor}         
\usepackage{amsmath, bm}
\usepackage{graphicx}
\usepackage{wrapfig}
\usepackage{booktabs}
\usepackage{multirow}
\usepackage{makecell}
\usepackage{amssymb}
\usepackage{mathtools}
\usepackage{amsthm}
\usepackage{multirow}
\usepackage{subfigure} 
\usepackage{float}
\usepackage{algorithm} 
\usepackage{algorithmic}
\usepackage{textcomp} 
\usepackage{listings} 
\usepackage{enumitem}
\usepackage{xcolor}
\usepackage{natbib}
\usepackage[utf8]{inputenc} 
\usepackage[T1]{fontenc}    
\usepackage{hyperref}       
\usepackage{url}            
\usepackage{booktabs}       
\usepackage{amsfonts}       
\usepackage{nicefrac}       
\usepackage{microtype}      
\usepackage{lipsum}
\usepackage{pifont}
\newcommand{\cmark}{\ding{51}}%
\newcommand{\xmark}{\ding{55}}%

\definecolor{lightblue}{rgb}{0.88, 0.96, 1}

\usepackage[normalem]{ulem}
\usepackage{colortbl}

\newif\ifdraft
\drafttrue 

\ifdraft
    
    \newcommand{\whr}[1]{\textcolor{magenta}{\sout{#1}}}
    \newcommand{\whc}[1]{\textcolor{magenta}{[WN: #1]}}
    \newcommand{\mlc}[1]{\textcolor{magenta}{[ML: #1}}
    \newcommand{\bsc}[1]{\textcolor{blue}{[BS: #1]}}
    \newcommand{\bck}[1]{\textcolor{green}{[BK: #1]}}
    \definecolor{av_comment}{RGB}{255, 128, 0}

\else
    
    \newcommand{\whr}[1]{}
    \newcommand{\whc}[1]{}
    \newcommand{\mlc}[1]{}
    \newcommand{\bsc}[1]{}
    \newcommand{\bck}[1]{}
\fi

\usepackage{lipsum}

\def\rr#1{\textcolor{black}{#1}}

\newcommand\blfootnote[1]{%
\begingroup
\renewcommand\thefootnote{}\footnote{#1}%
\addtocounter{footnote}{-1}%
\endgroup
}

\title{UniAudio: An Audio Foundation Model Toward Universal Audio Generation}

\author{Dongchao Yang$^{1*}$, Jinchuan Tian$^{2*}$, Xu Tan$^{3\dagger}$, Rongjie Huang$^4$, Songxiang Liu, Xuankai Chang$^2$, \\
\textbf{Jiatong Shi$^2$, Sheng Zhao$^3$, Jiang Bian$^3$, Zhou Zhao$^4$,  Xixin Wu$^1$,  Helen Meng$^{1\dagger}$}  \\
$^1$ The Chinese University of Hong Kong,
$^2$ Carnegie Mellon University, \\
$^3$ Microsoft Research Asia, $^4$ Zhejiang University \\ 
\texttt{dcyang@se.cuhk.edu.hk, jinchuat@andrew.cmu.edu} \\
}

\iclrfinalcopy 
\begin{document}

\maketitle
\begin{abstract}
\vspace{-10pt}

Large Language models (LLM) have demonstrated the capability to handle a variety of generative tasks. This paper presents the UniAudio system, which, unlike prior task-specific approaches, leverages LLM techniques to generate multiple types of audio (including speech, sounds, music, and singing) with given input conditions. 
UniAudio 1) first tokenizes all types of target audio along with other condition modalities, 2) concatenates source-target pair as a single sequence, and 3) performs next-token prediction using LLM. Also, a multi-scale Transformer model is proposed to handle the overly long sequences caused by the residual vector quantization-based neural codec in tokenization. Training of UniAudio is scaled up to 165K hours of audio and 1B parameters, based on all generative tasks, aiming to obtain sufficient prior knowledge not only in the intrinsic properties of audio but also the inter-relationship between audio and other modalities. Therefore, the trained UniAudio model has the potential to become a foundation model for universal audio generation: it shows strong capability in all trained tasks and can seamlessly support new audio generation tasks after simple fine-tuning. Experiments demonstrate that UniAudio achieves state-of-the-art or at least competitive results on most of the 11 audio generation tasks. Demo and code are released.\footnote{\url{https://uniaudio666.github.io/demo_UniAudio/}}
\blfootnote{* Equal contribution; $\dagger$ Corresponding author}.

\end{abstract}

\vspace{-10pt}
\section{Introduction}
\vspace{-10pt}
Audio generation is an important component of generative AI. Recently, the popularity of generative AI has induced increasingly emergent and varying needs in audio generation: audio is expected to be generated based on humans's demands, such as speech synthesis (TTS), voice conversion (VC), singing voice synthesis (SVS), text-to-sound, and text-to-music.
Prior works on audio generation tasks are commonly task-specific: their designs heavily leverage domain knowledge and their usage is restricted to fixed setups \citep{tts_survey, convtasnet, tse_survey, vc_survey, svs_survey}. 
Instead of taking care of each task independently, this work is an attempt to achieve universal audio generation, which intends to accomplish multiple audio generation tasks with only one unified model.
The universal audio generation model is expected to obtain sufficient prior knowledge in audio and related modalities, which has the potential to provide simple and 
effective solutions for the increasing needs of generating diverse types of audio.

The superiority of Large Languge Models (LLM) in text-generative tasks inspires a series of LLM-based models in audio generation \citep{valle, speartts, makeavoice, musiclm, borsos2023audiolm}.
    Among these works, LLM's capability in independent tasks has been extensively studied in tasks like text-to-speech (TTS) \citep{valle, speartts, makeavoice} and music generation \citep{musiclm,musicgen}, and achieves competitive performance.
However, LLM's ability to process multiple tasks with a unified model is less exploited in audio generation research: most existing LLM-based works are still designed for single tasks \citep{valle, speartts}.
We argue that achieving universality and versatility in audio generation through the LLM paradigm is promising but has not yet been comprehensively studied before this work.

Toward universal audio generation, this work presents UniAudio, which adopts LLM techniques and is able to generate multiple types of audio (speech, sounds, music, and singing) conditioned on various input modalities, such as phoneme sequences, textual descriptions, and audio itself.
The proposed UniAudio is mainly featured as follows:
First, all types of audio, along with all other input modalities, are tokenized as discrete sequences.
Specifically, a universal neural codec model is built to effectively tokenize audio regardless of the audio type, and other tokenizers are used to tokenize other different modalites. Then, UniAudio concatenates the source-target pair as a single sequence. Lastly, UniAudio performs next-token prediction using LLM.
The residual vector quantization \citep{soundstream} based on neural codecs is used in the tokenization process, resulting in overly long token sequences (one frame corresponding to multiple tokens) that cannot be processed efficiently by LLM. A multi-scale Transformer architecture is designed to reduce computational complexity by modeling the inter- and intra-frame correlation separately. Specifically, a global Transformer module is used to model the inter-frame correlation (\textit{e.g.} semantic level), and a local Transformer module is used to model the intra-frame correlation (\textit{e.g.} acoustic level). 

To demonstrate the scalability of UniAudio for new tasks, the building process of UniAudio takes two stages.
Firstly, the proposed UniAudio is trained on multiple audio generation tasks jointly, which allows the model to obtain sufficient prior knowledge not only of the intrinsic properties of audio but also of the interrelationship between audio and other input modalities. 
Secondly, through fine-tuning, the trained model can seamlessly support more unseen audio generation tasks.
Thus, UniAudio has the potential to become a foundation model for universal audio generation: it is able to continuously support emergent needs in audio generation.
Experimentally, our UniAudio supports 11 audio generation tasks: the training stage includes 7 audio generation tasks, while 4 tasks are further added in the fine-tuning stage.
The building process of UniAudio is scaled up to 165k hours of audio and 1B parameters.
Among the 11 tasks, UniAudio consistently obtains competitive performance in both objective and subjective evaluations. State-of-the-art results are even achieved on most of these tasks.
Further investigation suggests that training multiple tasks simultaneously in the training stage is mutually beneficial to each task involved. 
In addition, UniAudio can effectively adapt to new audio generation tasks and outperform task-specific models with a non-trivial gap.

To sum up, this work reveals that building universal audio generation models is necessary, promising, and beneficial. The main contributions of this work are summarized as follows: \\
(1) Toward universal audio generation, UniAudio is presented as a unified solution for 11 audio generation tasks.

(2) Per methodology, UniAudio provides novel approaches for (\romannumeral1) sequential representations of audio and other input modalities; (\romannumeral2) uniform formulation for LLM-based audio generation tasks; and (\romannumeral3) efficient model architecture specifically designed for audio generation.

(3) Per experiments, the overall performance of UniAudio is well validated, and the benefits of building a versatile audio generation model are verified by exhaustive experimental results.

(4) Demo and code are released, in the hope that UniAudio can become a foundation model that supports emergent audio generation in future research.
\vspace{-10pt}
\section{UniAudio}
\vspace{-10pt}
This section introduces the technical details of the proposed UniAudio. Section 2.1 explains how audio and other modalities are tokenized. Then, all considered audio generation tasks are uniformly formulated in Section 2.2. Subsequently, the multi-scale Transformer architecture is proposed in Section 2.3 to handle the overly long sequence challenge caused by the adoption of neural codecs.

\subsection{Tokenization}
\vspace{-5pt}
LLM are commonly used for sequential modeling, so audio and all other input modalities are tokenized before being processed.
These processes for each modality are completed by independent modules. All of these modules are fixed in the optimization of UniAudio or parameter-free.

\subsubsection{Audio} \label{sec:audio}

For all audio generation tasks considered in this work, audio, regardless of its types (speech, sounds, music, or singing), is the target to predict. 
Instead of modeling different types of audio separately, UniAudio intends to tokenize all types of audio as a single and unified modality (even though they commonly have distinct patterns, such as frequency span), which requires a model that is well-suited to mapping all audio types into a shared latent space.
Following \citet{valle, speartts}, neural codec models \citep{encodec, yang2023hifi, kumar2023highfidelity} are used in this work for audio tokenization.
An audio signal of duration $d$ with sample rate $f_s$ can be represented by a sequence $\mathbf{x} \in [-1,1]^{d*f_s}$. 
An audio neural codec intends to compress ${\mathbf{x}}$ and then recover it as $\mathbf{\hat{x}}$ using an encoder-decoder architecture with a quantization module:
\begin{equation}
    \mathbf{h} = \text{Encoder}(\mathbf{x}) \in \mathcal{R}^{T*L}; \quad 
    \mathbf{\hat{h}} = \text{Quantization}(\mathbf{h}); \quad 
    \mathbf{\hat{x}} = \text{Decoder}(\mathbf{\hat{h}})
\end{equation}
where $T$ denotes the number of audio frames after down-sampling in the encoder, and $L$ denotes the feature dimension of the encoder. The discrete representations of audio are the intermediate product of the quantization process. Given any frame of hidden output $\mathbf{h}_t$, the integer vector $\mathbf{z}_{t} = [z_t^1, ..., z_t^{n_q}]$ is generated by Residual Vector Quantization (RVQ) \citep{soundstream}, where $n_q$ denotes the number of vector quantization layers.
Iteratively, each element $z_t^k$ is the index among all pre-learned and fixed $k$-th level quantizer vectors $\{\mathbf{q}_k^*\}$ that has the smallest L2 distance to the residual between $\mathbf{h}_t$ and the sum of all previous chosen quantizer vectors $\{\mathbf{q}_{j}^{z_t^{j}}, j = 1,..., k-1\}$. With the discrete representation $\mathbf{z}_t$, $\mathbf{\hat{h}}_t$ is reconstructed as a close estimation of $\mathbf{h}_t$ that can be used to recover $\mathbf{x}_t$ with the decoder. 
\begin{equation}
\label{rvq}
    z_t^{k} = \argmin_{m} \text{Distance}(\mathbf{h}_t - \sum_{j=1}^{k-1} \mathbf{q}_{j}^{z_t^{j}}, \mathbf{q}_k^m);  \quad \mathbf{\hat{h}}_t = \sum_{j=1}^{n_q} \mathbf{q}_{j}^{z_t^{j}}; \quad 1 \leq k \leq n_q
\end{equation}
The discrete representation of all audio frames $\mathbf{z} \in \mathbf{Z}^{T\times n_q}$ is a matrix and needs to be converted into a sequence before being processed by LM: it is simply flattened as a sequence, in which every $n_q$ element for one frame is consecutive. Without specifically stated, we set $n_q=3$ in our experiments.
As the waveform can be recovered from $\mathbf{z}$ with a neural codec decoder, the rest of this paper mainly discusses how to predict the audio token sequence $\mathbf{z}$ using LLM techniques. 
\rr{As UniAudio intends to generate both speech and non-speech content, we build the codec model on our own and with broader data coverage. Details of our codec configuration is in Appendix \ref{appendix: codec}.}

\vspace{-5pt}
\subsubsection{Other modalities}
\vspace{-5pt}
Besides audio, other modalities considered in UniAudio also need to be represented as sequences. In addition, most of these sequences are transformed into discrete ones through tokenization.
The serialization and tokenization of these input modalities, along with their key features, are briefly summarized as below. 

\textbf{Phoneme:} Phonemes are the basic units of speech pronunciation in linguistics. 
Phoneme sequences have multiple sources: 
(1) when only text is available, phoneme sequence without duration information can be obtained by text-to-phoneme mapping using a pronunciation dictionary; 
(2) when only speech is available, phoneme sequence with duration information is obtained by beam search of the DNN-HMM system \citep{dnnhmm}; 
(3) when both text and speech are available, phoneme sequence with duration information is obtained by forced alignment of the DNN-HMM system
\footnote{CMUDict (http://www.speech.cs.cmu.edu/cgi-bin/cmudict) is adopted as the pronunciation dict; kaldi recipe (https://github.com/kaldi-asr/kaldi/tree/master/egs/librispeech/s5/local/chain/run\_tdnn.sh) is adopted to build the deep neural network-hidden Markov model (DNN-HMM) system.}.

\textbf{MIDI:} MIDI~\citep{m4singer} is widely used for singing voice synthesis tasks. F0 and duration information are included in the MIDI. We use the duration information to flatten the F0 sequence, so that the frame-level F0 sequence is obtained.

\textbf{Text:} Text acts as a effective carrier of human instructions in audio generation tasks \citep{yang2023instructtts,musicgen}. In this work, these textual instructions are represented as continuous embeddings derived from pre-trained text LLM \citep{raffel2020exploring}, as these embeddings contain rich textual semantics. 
Processing these continuous embeddings with LLM is further clarified in Section 2.3\footnote{The encoder of T5 (https://github.com/google-research/text-to-text-transfer-transformer) is used to extract the continuous text embeddings.}.

\textbf{Semantic Token:} The semantic tokens are derived from the continuous embeddings output by audio self-supervised learning (SSL) models. These continuous representations are highly informative and can be adopted in both speech understanding \citep{audiopalm} and generative tasks \citep{borsos2023audiolm}. Following \citet{makeavoice}, these continuous representations are tokenized by performing K-means clustering \citep{hubert} over these continuous representations. Since the continuous representations are frame-level, the semantic tokens also encode duration information\footnote{The 9-th layer hidden output of Hubert \citep{hubert} is adopted as the semantic token representations (https://github.com/facebookresearch/fairseq/hubert). The number of clusters for K-means is 500.}.

\subsection{Unified Task Formulation}

\begin{table*}[h]
    \centering
    \small
    \caption{Sequence formats of all tasks supported by UniAudio. 
    Text color represents modality. 
    black: audio; \textcolor{green}{green}: phoneme; \textcolor{blue}{blue}: MIDI; \textcolor{purple}{purple}: text; \textcolor{brown}{brown}: semantic token. 
    $\clubsuit$ means tasks that generate audio with deterministic length.
    $\diamondsuit$: means tasks that are only included in the fine-tuning stage. The speaker prompt is a 3-second speech and is used to represent the speaker identification.
    }
    \scalebox{0.85}{
    \begin{tabular}{lcc}
    \toprule    
    Task & Conditions & Audio Target \\
    \hline
    Text-to-Speech (TTS) \citep{valle} &  \textcolor{green}{phoneme}, speaker prompt & speech \\
    Voice Conversion (VC) $^\clubsuit$  \citep{lm-vc}  & \textcolor{brown}{semantic token}, speaker prompt & speech \\
    Speech Enhancement (SE) $^\clubsuit$ \citep{wang2023nadiffuse} & noisy speech & speech \\
    Target Speech Extraction (TSE) $^\clubsuit$ \citep{wang2018voicefilter} & mixed speech, speaker prompt & speech \\
    Singing Voice Synthesis (SVS) \citep{liu2022diffsinger} & \textcolor{green}{phoneme} (with duration), speaker prompt, \textcolor{blue}{MIDI} & singing  \\
    Text-to-Sound (Sound) \citep{diffsound} & \textcolor{purple}{textual description} & sounds \\
    Text-to-Music (Music) \citep{musiclm} & \textcolor{purple}{textual description} & music \\
    \hline
    Audio Edit (A-Edit) $^{\clubsuit\diamondsuit}$ \citep{wang2023audit} & \textcolor{purple}{textual description}, original sounds & sounds \\
    Speech dereverberation (SD) $^{\clubsuit\diamondsuit}$ \citep{wu2016reverberation} & reverberant speech & speech \\
    Instruct TTS (I-TTS)$^\diamondsuit$ \citep{guo2023prompttts} & \textcolor{green}{phoneme}, \textcolor{purple}{textual instruction} & speech \\
    Speech Edit (S-Edit) $^\diamondsuit$ \citep{tae2021editts}    &  \textcolor{green}{phoneme} (with duration), original speech  & speech\\
    \bottomrule
    \end{tabular} }
    \label{table:task_formats}
\end{table*}

For all tasks considered in UniAudio, the target audio is generated based on given conditions.
With the same target modality, i.e., audio, it is the conditions that define different audio generation tasks.
However, even with the variance in conditions, all tasks can still be uniformly formulated as sequential modeling tasks that can be processed by LLM: both the target audio and the conditions are first transformed as sub-sequences and spliced as [\textit{conditions}, \textit{target}] sequences to be processed.

UniAudio supports 11 audio generation tasks in total.
The sequential formats of each task are defined in Table \ref{table:task_formats}, in which the sub-sequences of all modalities are derived as in Section 2.1. However, due to the unique configurations of each task, some of the condition sub-sequences are subject to task-specific pre-processing operations during the tokenization. 
For audio, these operations are mainly for data corruption, such as adding noise, reverberation, and speech mixed with other speakers in the raw audio before tokenization.
For phoneme and semantic tokens, duration information is reserved by default but can also be removed. For singing voice synthesis and speech edit tasks, the duration information of phoneme is used. For TTS and I-TTS tasks, the duration information is not used. For MIDI, the duration information is used repeat the F0 sequence.
For text embeddings, no operations are applied in this work.

To avoid ambiguity, some special discrete tokens (enclosed by \texttt{<>}) are inserted to indicate (1) the start and end of the whole sequence; (2) the start and end of each sub-sequence of a certain modality; and (3) the task identifier. For example, for a text-to-sound task sequence that generates target audio based on textual description, the whole sequence is like:
\textit{<start> <sound\_task> <text\_start> text\_sequence <text\_end> <audio\_start> audio\_sequence <audio\_end> <end>}.

\vspace{-5pt}
\subsection{Multi-Scale Transformer}
\vspace{-5pt}

\begin{figure}[t]
    \centering
    \vspace{-5mm}
    \includegraphics[width=\textwidth]{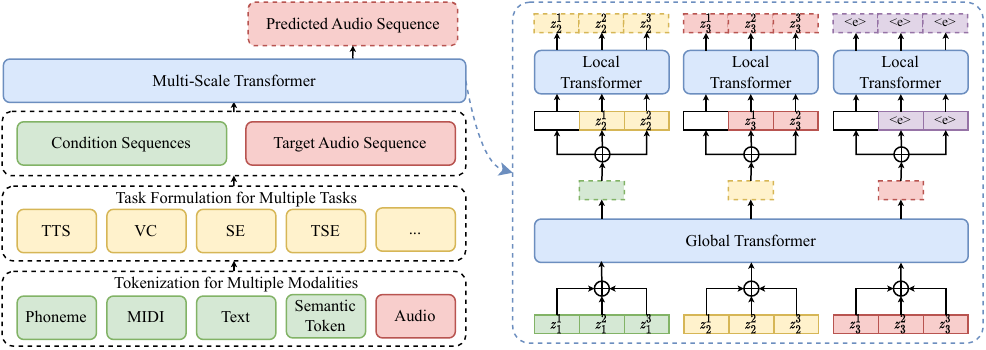}
    \vspace{-8mm}
    \caption{Overview of UniAudio (left) and multi-scale Transformer architecture (right). <e> represent the end of the sequence. $z_t^k$ denotes the k-th audio token at t-th frame.} 
     \vspace{-6mm}
    \label{fig:multi_scale_transformer}
\end{figure}

\label{chapter:model_arch}
Previous work on LLM-based audio generation \citep{musicgen} advocates to modeling the discrete audio tokens as flattened sequences. 
If so, these sequences are processed in the length of $T \times n_q$, which is highly challenging considering the quadratic space complexity of Transformer \citep{vaswani2017attention} with respect to the lengths.
Inspired by \citet{yu2023megabyte}, a multi-scale Transformer architecture is specifically designed for discrete audio sequences, which is a hierarchical model that processes the inter- and intra-frame correlation by global and local Transformer modules separately. 
An overview of the proposed architecture is in Figure \ref{fig:multi_scale_transformer}.
Instead of processing the whole flattened sequence token-by-token like prior works \citep{speartts}, the multi-scale transformer considers patches (i.e., every consecutive $n_q$ token) as the global modeling units and then handles the tokens within each patch locally. Note that both the global and local Transformers are causal.

For audio token sequences, each patch accounts for $n_q$ consecutive audio tokens that exactly represent one audio frame.
First, as suggested in Equation \ref{rvq}, regardless of the exact choices of each quantization vector $\mathbf{q}_*^{z_t^*}$, it is the summed quantization vector $\mathbf{\hat{h}}_t$ that is used to represent the audio frame. 
Thus, in the embedding stage, each patch (a.k.a., frame) is represented by the summed vector of the corresponding embeddings before entering the global Transformer. 
Second, the global Transformer is to predict audio frame-by-frame: to predict the frame $\mathbf{x}_t$, it outputs the continuous representations that include frame $\mathbf{x}_{t-1}$ and all previous content. These continuous representations will be further processed by the local Transformer.
Third, also as in Equation \ref{rvq}, given the hidden representation $\mathbf{h}_t$, the acquisition of $\mathbf{z}_t$ is independent of any hidden output other than $\mathbf{h}_t$.
Inspired by this, it is reasonable to predict the discrete tokens for frame $\mathbf{x}_t$, a.k.a., patch $\mathbf{z}_t$, only with the hidden output of global Transformer corresponding to frame $\mathbf{x}_{t-1}$. 
To be more detailed, as the acquisition of each token $z_t^k$ is auto-regressively dependent on its prior tokens $\{z_t^{j} | j < k\}$, a local Transformer is adopted to predict the patch sequence $\mathbf{z}_t$ in auto-regressive style. During this process, the corresponding vector output by the global transformer acts as a patch-level context, which is linearly transformed and then added to the embedded results of each token $z_t^k$. 

The proposed multi-scale Transformer architecture is also compatible with discrete and continuous sequences besides audio.
For all discrete tokens except audio (phoneme, semantic, MIDI and special tokens), each token has independent semantics and thus should account for one patch. So these discrete tokens repeat for $n_q$ times to fill each patch.
The continuous text embeddings are also repeated for $n_q$ times for the same purpose. Additionally, their embedding process is replaced by a linear transformation while their predicting targets for local Transformer are consecutive special tokens \textit{<continuous\_token>}.

The design of the proposed multi-scale Transformer can effectively reduce computational complexity.
First, the equivalent sequence length for the global Transformer is reduced from $T \times n_q$ to $T$, which makes the global modeling cost independent to $n_q$ and thus the adoption of a larger $n_q$ becomes feasible.
Second, the intra-patch computation to generate the discrete tokens for each frame is offloaded to the local Transformer. The computation on the local transformer is comparatively light since it only processes the very short sequence (fixed to the length of $n_q$) and empirically has fewer parameters than the global Transformer by design.

\vspace{-5pt}
\section{Experiments}
This section first introduces the experimental setup in Section 3.1. The results for the training stage and the fine-tuning stage are presented in Section 3.2 and 3.3 respectively. 
Ablation studies are presented in Section 3.4.

\subsection{Experimental Setup}

\textbf{Data and Model:} \rr{UniAudio is built on labeled datasets.} Specifically, 12 datasets are adopted in this work, all of which are publicly available. The overall audio volume is 165K hours.
Detailed data statistics and their adoption for each task are in Appendix \ref{appendix:data}.
Discrete tokens from all modalities form a joint vocabulary of size 4212, including all special tokens.
Vanilla Transformer decoder layers with causality are consistently adopted in global and local Transformer. The overall parameter budget is roughly 1B. Detailed model configuration is in Appendix \ref{appendix: model config}. 
Existing neural codec models are sub-optimal for universal audio generation, mainly due to data coverage. An improved neural codec model is then built with fewer quantization levels $n_q$, smaller frame-per-second rate, higher quality, and wider coverage (see Appendix \ref{appendix: codec}).

\textbf{Training and Inference: }
The training stage includes 7 tasks while 4 new tasks are added in the fine-tuning stage. 
Table \ref{table:task_formats} specifies the tasks for fine-tuning only.
Both the training and fine-tuning are completed with 16 AMD MI200-64G GPUs. The detailed configuration of optimization is in Appendix \ref{appendix: optimization}. To retain the performance of previous tasks during fine-tuning, following \citet{xlsr}, the training data are re-sampled with respect to tasks with $\alpha=0.05$. Top-k sampling is adopted consistently for inference, in which $k$ and the temperature are set to 30 and 0.8, respectively. 
As the global Transformer does not directly predict tokens, the sampling process only happens in the local Transformer inference.

\textbf{Evaluation: } For evaluation, most tasks are evaluated using both objective and subjective metrics \footnote {Following the setting of DiffSinger \citep{liu2022diffsinger}, SVS tasks don't report the objective results}. 
Generally, for objective evaluation,
Word Error Rate (WER) is used to evaluate the intelligibility of generated speech; Similarity Score (SIM) is for similarity in terms of speaker identity\footnote{WER and SIM evaluation models follow \citet{valle}}; Perceptual Evaluation of Speech Quality (PESQ), VISQOL\footnote{\url{https://github.com/google/visqol}}, DNSMOS \footnote{\url{https://github.com/microsoft/DNS-Challenge/tree/master/DNSMOS}} and Mel Cepstral Distortion (MCD) are signal-level quality metrics derived from human auditory research; 
Following \citep{musicgen}, Fréchet Audio Distance (FAD), Kullback-Leiber (KL) Divergence, and Fréchet Distance (FD) are for audio fidelity and audio similarity; For subjective evaluation, MOS and SMOS are adopted to provide human-centric judgment for speech and sing related tasks. For text-to-sound and text-to-music tasks, we use overall quality (OVL), and relevance to the text input (REL) \citep{musicgen}.
Note all subjective results are obtained from Amazon Mechanical Turk\footnote{\url{https://www.mturk.com/}} for fair comparison. Appendix \ref{appendix: evaluation metric} shows details of the subjective evaluation process.

\subsection{The Results of 7 generative tasks in the training stage}
\label{sec_pretraining}

\begin{table}[t]
    \centering
    \caption{Performance evaluation for UniAudio and selected prior works in the training stage
    }
    \scalebox{0.79}{
    \begin{tabular}{cccccc}
    \toprule
    \multirow{2}{*}{Task} & \multirow{2}{*}{Model} &  \multicolumn{2}{c}{Objective Evaluation}  & \multicolumn{2}{c}{Subjective Evaluation} \\  
    && Metrics & Results & Metrics & Results \\
    \midrule
    \multirowcell{2}{Text-to-Speech} & \citet{naturalspeech2} & \multirow{2}{*}{SIM$(\uparrow)$ / WER$(\downarrow)$} & 0.62 / 2.3 & \multirowcell{2}{MOS$(\uparrow)$ \\/ SMOS$(\uparrow$)} & \textbf{3.83$\pm$0.10} / 3.11$\pm$0.10 \\ 
    & UniAudio & & \textbf{0.71 / 2.0} & & 3.81$\pm$0.07 / \textbf{3.56$\pm$0.10} \\
    \hline
    \multirowcell{2}{Voice \\ Conversion} & \citet{lm-vc} & \multirowcell{2}{SIM$(\uparrow)$ / WER$(\downarrow)$} & 0.82 / 4.9 & \multirowcell{2}{MOS$(\uparrow)$ \\/ SMOS$(\uparrow$)} & 3.41$\pm$0.08 / 3.17$\pm$0.09 \\ 
    & UniAudio & & \textbf{0.87 / 4.8} & & \textbf{3.54$\pm$0.07 / 3.56$\pm$0.07} \\
    \hline
    \multirowcell{2}{Speech \\ Enhancement} & \citet{richter2023speech} & \multirowcell{2}{PESQ$(\uparrow)$ / VISQOL$(\uparrow)$ \\ / DNSMOS$(\uparrow)$}& \textbf{3.21 / 2.72} / 3.29 & \multirow{2}{*}{ MOS$(\uparrow$)} & 3.56$\pm$0.08 \\ 
    & UniAudio &  & 2.63 / 2.44 / \textbf{3.66} & & \textbf{3.68$\pm$0.07} \\
    \hline
    \multirowcell{2}{Target Speaker \\ Extraction} & \citet{wang2018voicefilter} & \multirowcell{2}{PESQ$(\uparrow)$ / VISQOL$(\uparrow)$ \\ / DNSMOS$(\uparrow)$} & \textbf{2.41 / 2.36} / 3.35 & \multirow{2}{*}{ MOS$(\uparrow$)} & 3.43$\pm$0.09 \\ 
    & UniAudio &  & 1.88 / 1.68 / \textbf{3.96} & & \textbf{3.72$\pm$0.06} \\
    \hline
    \multirowcell{2}{Singing Voice \\ Synthesis} & \citet{liu2022diffsinger} & \multirow{2}{*}{-} & \multirow{2}{*}{-} & \multirowcell{2}{MOS$(\uparrow$) \\/ SMOS$(\uparrow$)} & 3.94$\pm$0.02 / \textbf{4.05$\pm$0.06} \\ 
    & UniAudio &  &  & & {4.08$\pm$0.04} / 4.04$\pm$0.05 \\
    \hline
    \multirowcell{2}{Text-to-Sound} & \citet{liu2023audioldm} & \multirow{2}{*}{FAD $(\downarrow)$ / KL $(\downarrow)$} & 4.93 / 2.6 & \multirowcell{2}{OVL $(\uparrow)$ \\/ REL $(\uparrow)$} & 61.0$\pm$1.9 / 65.7$\pm$1.8 \\ 
    & UniAudio &  & \textbf{3.12 / 2.6} & & \textbf{61.9$\pm$1.9 / 66.1$\pm$1.5} \\
    \hline
    \multirowcell{2}{Text-to-Music} & \citet{musicgen} & \multirow{2}{*}{FAD $(\downarrow)$ / KL $(\downarrow)$} & 4.52 / \textbf{1.4} & \multirowcell{2}{OVL $(\uparrow)$ \\/ REL $(\uparrow)$} & \textbf{73.3$\pm$1.5 / 71.3$\pm$1.7}
    \\
    & UniAudio &  & \textbf{3.65} / 1.9 & & 67.9$\pm$1.7 / 70.0$\pm$1.5 \\
    \bottomrule 
    \vspace{-25pt}
    \end{tabular}}
    \label{tab:all_results}
\end{table}

\rr{This section presents the overall evaluation results of the proposed UniAudio model over all 7 audio generation tasks during the training stage. A comprehensive comparison is conducted between UniAuduio and multiple prior works on each task, including not only the LM-based methods but also the diffusion model-based methods as well as other conventional audio generation methods. The detailed comparison is presented in Appendix \ref{appendix:exp}. We selected one of the most advanced prior work in each task and present the results in Table \ref{tab:all_results}.}

\rr{As suggested in Table \ref{tab:all_results}, UniAudio is a versatile system that can handle all 7 audio generation tasks together and achieve competitive performance. 
Per subjective evaluation, UniAudio surpasses the baselines in 3 out of 6 tasks (TTS, VC, Sound); per objective evaluation, it achieves better results on 5 out of the 7 tasks except SVS and Music. 
We also find UniAudio under-perform on several metrics.
UniAudio's subjective performance for SE and TSE is less competitive compared with its competitors, which is also observed in previous literature \citep{tokensplit} that the signal-level evaluation metrics may not be suitable for LM-based generative methods.
UniAudio cannot surpass the selected competitor \citep{musicgen} in the Text-to-Music task. We note that \citep{musicgen} is built with more private labeled data than our UniAudio.}


\vspace{-5pt}
\subsection{The Results of 4 generative tasks in the fine-tuning stage}

\begin{table}[h]
    \centering
    \vspace{-18pt}
    \caption{Performance evaluation for UniAudio and selected prior works in the fine-tuning stage}
    \scalebox{0.75}{
    \begin{tabular}{cccccc}
    \toprule
    \multirow{2}{*}{Task} & \multirow{2}{*}{Model} &  \multicolumn{2}{c}{Evaluation} \\  
    && Metrics & Results  \\
    \midrule
    \multirowcell{2}{Audio Edit} & AUDIT \citep{wang2023audit} & \multirow{2}{*}{FD $(\downarrow)$ / KL $(\downarrow)$} & 20.78 / 0.86  \\ 
     & UniAudio &  & \textbf{17.78 / 0.77}  \\
    \hline 
    \multirowcell{2}{Speech Dereverb.} & SGMSE+ \cite{richter2023speech} & \multirow{2}{*}{PESQ$(\uparrow)$ / DNSMOS$(\uparrow$)} & \textbf{2.87} / 3.42  \\ 
    & UniAudio  &  & 2.13 / \textbf{3.51}   \\
    \hline 
    \multirowcell{2}{Instructed TTS} & GroundTruth & \multirowcell{2}{MOS$(\uparrow)$ / SMOS$(\uparrow$)} & \textbf{3.77$\pm$0.07} / \textbf{3.85$\pm$0.08} \\ 
    & UniAudio   & & 3.61$\pm$0.09 / 3.71$\pm$0.09  \\
    \hline 
    \multirowcell{2}{Speech Edit} & TTS system regeneration  & \multirow{2}{*}{MCD($\downarrow$) / MOS$(\uparrow)$} & 6.98 / 3.69$\pm$0.08 \\ 
       & UniAudio & & \textbf{5.12} / \textbf{3.82$\pm$0.06} \\
    \bottomrule
    \end{tabular} }
    \label{tab:fine-tune} 
\end{table}


\rr{As UniAudio is designed to continuously support new audio generation tasks, this section reports UniAudio's performance on unseen tasks. The model is obtained by fine-tuning over 4 new tasks jointly and the results are presented in Table \ref{tab:fine-tune}. Similar to section \ref{sec_pretraining}, for each task, we compare UniAudio's performance with one selected prior work and report the detailed results in Appendix \ref{appendix:exp}.}

\rr{
As shown in Table \ref{tab:fine-tune}, the fine-tuned UniAudio model surpasses its baselines in audio edit and speech dereverberation and is approaching the ground-truth quality in the Instructed TTS task. For speech editing, UniAudio shows considerable improvement compared to generating the whole sentence. 
}

\vspace{-5pt}
\subsection{Ablation Study}
\vspace{-5pt}
\subsubsection{Benefit of building unified audio generation model}
\rr{
To further validate our claim that building a unified model for all 11 audio generation tasks is promising and beneficial, more ablation studies are conducted.
In Appendix \ref{ablation: multi-task}, we demonstrate that the joint-trained UniAudio model consistently outperforms the models that are trained for each specific task\footnote{Note the task-specific models are built with the corresponding subset of the training data.}, regardless they are included in the training stage or the fine-tuning stage. In Appendix \ref{ablation: fine-tune}, we additionally validate that fine-tuning over the 4 new audio generation tasks does not affect UniAudio's performance on the original 7 tasks. In Appendix \ref{ablation: data-quantify}, we observe that UniAudio can consistently benefit from increased training data volume of each task, which provides another reason to build universal audio generation models: these models are easier to scale up as the data collection is more feasible. We provide more discussion in Appendix \ref{why work} about the effectiveness of building a universal audio generation model.
}

\vspace{-5pt}
\subsubsection{The effectiveness of multi-scale transformer model}
\vspace{-5pt}
\begin{figure*}[h]
    \centering
    \includegraphics[width=0.85\textwidth]{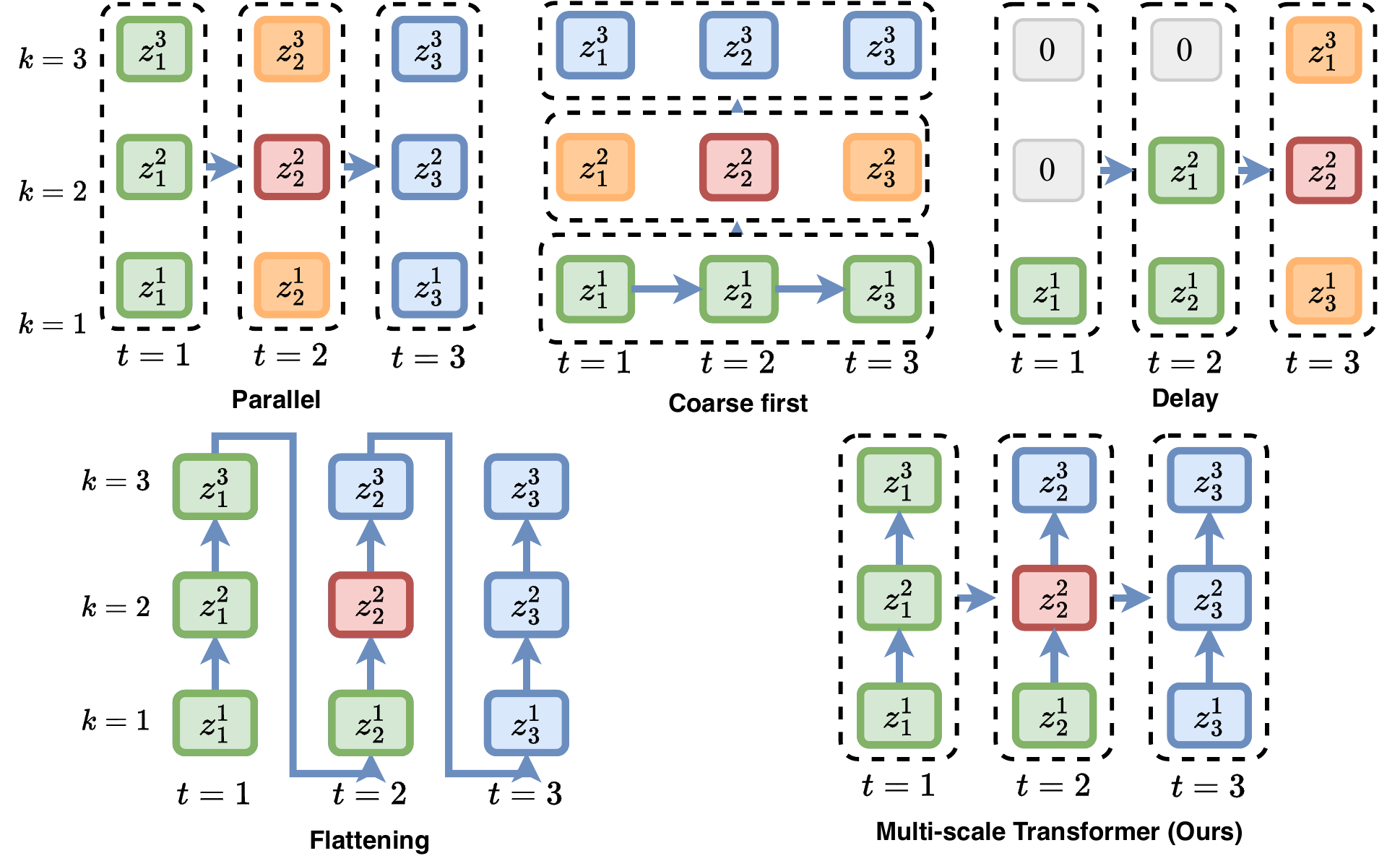}
    \vspace{-4mm}
    \caption{Order of token prediction for 4 representative methods in audio generation \citep{musicgen} and the proposed multi-scale Transformer. Assume $n_q=3$ and $T=3$. Current token prediction (red) is conditioned on prior tokens (in green). Tokens in orange are concurrently predicted with the current token. 0 is a special token indicating empty positions in the delay prediction.}
    \label{fig:predict_order}
\end{figure*}

\rr{
As in section \ref{chapter:model_arch}, the adoption of neural codecs has become a popular choice of LLM-based audio generation but causes an overly long sequence issue that needs further consideration. This section compares the proposed multi-scale Transformer with four representative approaches in this field: Flattening Prediction (\textit{e.g.} SPEARTTS \citep{speartts}), Coarse first prediction (\textit{e.g.} VALL-E \citep{valle}), Parallel prediction (\textit{e.g.} AudioGen \citep{kreuk2022audiogen}), and Delay prediction (\textit{e.g.} MusicGen \citep{musicgen}). Figure \ref{fig:predict_order} illustrates the prediction order of these five architectures. Experiments are conducted on text-to-speech and text-to-music tasks and the results are reported in Table \ref{tab:structure-comparison} and \ref{tab:ablation-structure-music} respectively \footnote{Results are based on unofficial implementations.}.}

\begin{table}[t] \label{tab:structure comparison}
    \centering
    \small
    \vspace{-15pt}
    \caption{Model comparison among Coarse first, Flattening, Parallel, delay prediction, and multi-scale Transformer. Experiments were conducted on the LibriTTS. GPU memory and training time are obtained by a 20-second audio (average of 100 trials). All models have a similar parameter budget.
    }
    \scalebox{0.85}{
    \begin{tabular}{lcccccc}
    \toprule
    Structure & $n_q$  & MOS ($\uparrow$) & MCD ($\downarrow$) & GPU Mem. (GB) & Time (s) / Iter.   \\
    \midrule
    Coarse first   & 8  & 3.48$\pm$0.05 & 7.37 & 18.7 & 0.58   \\
    Parallel    & 3   & 3.14$\pm$0.07 & 7.89 & 13.56 & 0.53    \\
    Delay    & 3  & 3.48$\pm$0.05 & 6.95 & 13.65 & 0.59    \\
    Flattening  & 3  &3.80$\pm$0.09 & 6.56 & 36.7  & 1.63   \\
    
    Multi-Scale Transformer (ours)  & 3 & 3.77$\pm$0.05 
 & 6.52 & 19.4 & 0.73  \\
    Multi-Scale Transformer (ours)  & 8 
    &  3.84$\pm$0.06 & 6.27 & 24.0 & 1.10  \\
    \bottomrule
    \end{tabular}}
    \vspace{-10pt}
    \label{tab:structure-comparison}
\end{table}

\rr{
\textbf{Auto-Regression and Performance:} 
Among all 4 baselines aforementioned, \citet{musicgen} claims that the flattening method provides the best audio generation quality. they further claim that the superior performance of flattening prediction is mainly attributed to the auto-regressive property; the other three methods do not reserve this property as the concurrent prediction is introduced (see Fig. \ref{fig:predict_order}). 
Under the scenario of codec adoption, we reinterpret the auto-regressive property as: current token prediction is based on all tokens of previous frames and the previous tokens within the current frame, or formally, the prediction of the current token $z_t^k$ is based on tokens: $\{z_{t'}^{k'} | t' < t\} \cup \{z_{t'}^{k'} | t' = t,  k' < k\}$. With this definition, we claim that the proposed multi-scale transformer is also auto-regressive.}

\rr{
Aligned with \citet{musicgen}, our experiments also validate the importance of the auto-regressive property. As in Table \ref{tab:structure-comparison} and \ref{tab:ablation-structure-music}, flattening prediction brings better generation quality than parallel, coarse first, and delay prediction. Additionally, with the same auto-regressive property, our proposed multi-scale transformer achieves a comparable performance with flattening prediction in terms of generation quality, which, again, validates the importance of auto-regression.}

\rr{
\textbf{Efficiency: } Besides generation quality, efficiency is a major concern of audio generation. Although with the auto-regressive property, the flattening prediction is sub-optimal in terms of efficiency: the modeling is based on the $T \times n_q$ long sequence, which has a space complexity of $O((T*n_q)^2)$ in self-attention. As increasing $n_q$ gives higher reconstruction quality at the cost of longer sequences and more computation, this issue becomes more severe when a larger $n_q$ is adopted. Since the sequence length grows proportionally with $n_q$, we experimentally find it difficult to train with $n_q \ge 4$. By contrast, the proposed multi-scale Transformer distributes the inter- and intra-frame modeling to the global and local sub-modules respectively, which thus alleviates the space complexity to $O(T*^2)$. Finally, without the requirement of auto-regression, methods like parallel, coarse first, and delay predictions achieve better efficiency due to the adoption of concurrent predictions. Since the space complexity is independent to $n_q$, training a larger $n_q$ with the multi-scale transformer is then feasible.}

\rr{
Experimentally, the proposed multi-scale transformer considerably reduces the time and memory cost compared with the flatting prediction. It still costs more time and memory compared with the other three baselines.
}

\rr{
Based on the observations above, we claim that the proposed multi-scale transformer is an auto-regressive architecture that achieves a better trade-off between generation quality and efficiency.
}


\begin{table}[t] 
    \centering
    \small
    \caption{The ablation study to explore the effectiveness of our proposed multi-scale transformer. Experiments were conducted on text-to-music tasks with the Million Song dataset.}
    \label{tab:ablation-structure-music}
    \begin{tabular}{l|ccc|cc}
    \toprule    
    Structure & $n_q$ & FAD ($\downarrow$)  & KL ($\downarrow$)   & OVL. ($\uparrow$)   & REL. ($\uparrow$)  \\
    \midrule
    Parallel   & 3 & 6.92   &  2.36  & 60.4$\pm$2.3  & 61.3$\pm$1.5   \\
    Delay & 3  & 6.07   & 2.23  & 62.8$\pm$1.9   & 63.9$\pm$1.6   \\
    Flatten & 3   & {5.18}    & 1.83   & {64.8$\pm$1.8} &  65.2$\pm$2.0  \\
    Multi-Scale Transformer (ours) & 3  & 5.24   &  {1.80}   & 64.4$\pm$2.1   & {66.2$\pm$2.4} \\
    \bottomrule 
    \end{tabular}
    \vspace{-10pt}
\end{table}

\vspace{-12pt}
\section{Related Works}
\vspace{-12pt}
This work is an attempt to achieve universal audio generation through LLM-based techniques.
\rr{There is a long research history for many audio generation tasks. Conventionally}, the design of these tasks heavily leverages the domain knowledge of each specific task, and their workflows are distinctive from each other:
For tasks like TTS, SE, TSE, TT-Music, VC, S-Edit, SD, SVS,
(1) their neural network architectures are based on Transformer \citep{fastspeech2} or others \citep{oord2016wavenet,convtasnet}; 
(2) their training objectives can be either in time-domain \citep{convtasnet}, frequency-domain \citep{yu2017permutation} or others \citep{gu2021complex,naturalspeech2}; 
(3) their designs are inspired by and derived from linguistics and phonetics \citep{zen2013statistical}, signal processing \citep{griffin1984signal}, auditory perception \citep{shadle2001prospects} and machine learning \citep{wang2016first} research, etc; 
(4) they use different generative models, such as diffusion model \citep{naturalspeech2,wang2023nadiffuse}, flow \citep{voicebox}, Seq2Seq \citep{fastspeech2,ppg-vc}. 



The prosperity of LLM techniques \citep{radford2019language,gpt4} significantly promotes progress in audio generation research in several directions.
First, the large language models, along with the prompt methods, inspired multiple emergent audio generation tasks that are based on textual instruction or descriptions from humans, such as Instruct-TTS \citep{yang2023instructtts}, Text-to-sound \citep{kreuk2022audiogen,make-an-audio} and text-to-music \cite{musicgen,musiclm}.
Second, besides the text, audio can also be tokenized as discrete sequences \citep{soundstream, encodec, kumar2023highfidelity} that can be further processed by LMs. LM-based audio generative models then show superior capability in generalization towards unseen speakers \citep{valle}, low resources \citep{speartts} and multilingual \citep{vallex} scenarios. These methods also achieve state-of-the-art results in overall performance within their own scopes.
\rr{Finally, the LM-like model can be further combined with existing generative models (e.g., diffusion models \cite{rombach2022high}) to obtain improved generation quality.}

It is laborious to handle each audio generation task case-by-case, especially when considering the data shortage as well as the emergent and varying needs in this area. Alternatively, building a universal audio generation model is a promising and practical paradigm. 
\rr{Given the rapid progress in audio generation research, recent designs of audio generation, including LM-based ones, tend to support multiple audio generation tasks simultaneously. Some pioneer works \citep{speechx, voicebox,naturalspeech2,audioldm2,megatts} clearly consider supporting multiple tasks as a key strength; the designs of other prior works \citep{borsos2023audiolm,speartts,naturalspeech2} do show the potential to generate audio in a broader sense than what they originally claim. Following these pioneering research works, UniAudio supports an extended coverage of 11 audio generation tasks in a unified LM-based model. 
}


\vspace{-12pt}
\section{Limitation}
\vspace{-12pt}
Not all known audio generation tasks are included in the proposed UniAudio, \rr{such as noise removal, noise speech edit \citep{speechx} and speech-to-speech translation \citep{audiopalm,barrault2023seamlessm4t}.}
All new tasks added in fine-tuning are formulated with the known modalities in the training stage;
\rr{Introducing new modalities during fine-tuning is unexplored in this work. Current UniAudio considers neither unlabeled data nor domain-specific foundation models, which can possibly further improve the overall performance. The samples generated by UniAudio are not guaranteed in quality and may contain errors.}

\vspace{-10pt}
\section{Conclusion}
\vspace{-10pt}
To handle the emergent and varying needs in audio generation, this work is an attempt to achieve universal audio generation. 
\rr{UniAudio is proposed as a unified LM-based generative model that supports 11 different audio generation tasks.}
In experiments, the proposed UniAudio provides competitive performance on all 11 tasks.
It also empirically demonstrates the capability of continuously integrating unseen audio generation tasks. 
Demo and code are released, in the hope that UniAudio can become a foundation model for universal audio generation in further research.

\section{Ethical Statement}
We are delving into the revolutionary field of generating diverse audio using large language model techniques.
We find ourselves at the confluence of innovation and responsibility. It is imperative to acknowledge the ethical dimensions of our work and ensure that our contributions are employed for the betterment of society.

\textbf{Being Open: } As we advance in this domain, it's crucial to ensure that the benefits of this technology are widespread and not limited to a privileged few. Our code is released publicly along with this submission to ensure equal access for each person. All experiments are based on open-accessible datasets that allow research-oriented comparison and reproduction.

\textbf{Avoid Misuse: } While our model can produce a myriad of audio content ranging from music to speech, there's potential for misuse in the generation of misinformation, deepfake audio, or any harmful content.
We advocate for adopting our code and model responsibly, with full respect to individual privacy and observance of regulations.
Concerning the potential misuse of our model, checkpoints will not be released.

\bibliography{iclr2024_conference}
\bibliographystyle{iclr2024_conference}

\newpage
\appendix
\begin{center}{\bf {\LARGE Appendices} }
\end{center}

\section{Experimental Setup}
This appendix describes experimental setups in detail, including data statistics, model architecture and optimization strategy.

\subsection{Data Description} \label{appendix:data}
12 public datasets are adopted in this work for training. Besides, several test sets are additionally used only for zero-shot evaluation. The statistics of these datasets are in Table \ref{tab:data_statistics}. Datasets adoption for each task is described in Table \ref{tab:data_task_assign}. Note some datasets are adopted by more than one task.

\begin{table}[h]
    \centering
    \small
    \caption{Data statistics}
    \scalebox{0.9}{
    \begin{tabular}{lccc}
    \toprule
    Dataset & Type & Annotation & Volume (hrs) \\
    \midrule
    \multicolumn{4}{l}{Training} \\ 
    \midrule
        LibriLight \citep{librilight} &  speech & -  & 60k \\
        LibriTTS \citep{libritts} & speech       & text  & 1k \\ 
        MLS \citep{pratap2020mls} & speech & -  & 20k \\
        AudioSet \citep{audioset} & sound  & -  & 5.8k    \\
        AudioCaps \citep{audiocaps} & sound & text description  & 500 \\
        WavCaps \citep{mei2023wavcaps} & sound & text description  & 7k \\
        Million Song Dataset \citep{msd} & music & text description  & 7k \\
        OpenCPOP \citep{opencpop}  & singing & text, MIDI  & 5.2 \\
        OpenSinger \citep{opensinger}  & singing & text, MIDI & 50 \\
        AISHELL3 \citep{shi2020aishell}  & speech & text  & 85 \\
        PromptSpeech \citep{guo2023prompttts} & speech & text, instruction  & 200 \\
        openSLR26,openSLR28 \citep{ko2017study} & room impulse response  & -  & 100 \\
    \midrule
    \multicolumn{4}{l}{Test} \\ 
    \midrule
    LibriSpeech test-clean \cite{librispeech} & speech       & text  & 8 \\ 
    VCTK \citep{vctk} & speech       & text  & 50 \\
    TUT2017 Task1 \citep{tut2017} & Noise & -  & 10 \\
    Cloth \citep{drossos2020clotho} & Sound &  text description  & 3 \\
    MusicCaps \citep{musiclm} & Music  &   text description  & 15 \\
    M4Singer\citep{m4singer} & singing   & text, MIDI  & 1 \\
    \bottomrule
    \end{tabular} } 
    \label{tab:data_statistics}
\end{table}

\begin{table}[h]
    \centering
    \small
    \caption{Dataset adoption of all tasks}
    \scalebox{0.9}{
    \begin{tabular}{lccc}
    \toprule
    Task   & Training dataset & Test set &  Train Volume (hrs) \\
    \midrule
    \multicolumn{4}{l}{Training Stage} \\ 
    \midrule
    TTS  & Librilight& {LibriSpeech} clean-test & 60k  \\
    VC   & Librilight & VCTK & 60k \\
    SE   & MLS, Audioset  & TUT2017 Task1, VCTK & 20k \\
    TSE  & MLS  & Libri2Mix test set & 10k \\
    Sound & AudioCaps, WavCaps & Cloth test set & 7k \\
    Music & MSD  & MusicCaps & 7k \\
    Singing  & OpenCPOP, OPenSinger, AISHEELL-3 & M4Singer test set &  150 \\
    \midrule
    \multicolumn{4}{l}{Fine-Tuning Stage} \\ 
    \midrule
    I-TTS & PromptSpeech & PromptSpeech test set & 200 \\
    Speech dereverberation & LibriTTS, openSLR26, openSLR28 & LibriTTS test set & 100 \\
    Speech edit & LibriTTS & LibriTTS test set & 100 \\
    Audio edit & AudioCaps, WavCaps & AudioCaps test set & 500 \\
    \midrule
    Sum & - & - & 166k \\ 
    \bottomrule 
    \end{tabular} }
    \label{tab:data_task_assign}
\end{table}

\subsection{Model Configuration} \label{appendix: model config}
The model configuration of the proposed multi-scale Transformer is described in Table \ref{tab:model_config}.
\begin{table}[h]
    \centering
    \small
    \caption{Model configuration (with $n_q=3$)}
    \begin{tabular}{lcc}
    \toprule
    Hyper-parameter & Global Transforemr & Local Transformer \\
    \midrule
    \#layer & 24 & 8 \\
    \#Attention dim & 1536 & 1536 \\
    \#Attention head & 12 & 12 \\
    \#Feed-Forward dim & 6144 & 6144 \\
    \#Params (M) & 744 &  238 \\
    Max context length (in \#tokens) & 3,000 & 3 \\
    Causality & Yes & Yes \\
    \bottomrule
    \end{tabular}
    \label{tab:model_config}
\end{table}

\subsection{Optimization} \label{appendix: optimization}
The optimization configurations adopted in both the training and fine-tuning stages are presented in Table \ref{tab:optim_config}

\begin{table}[h]
    \centering
    \small
    \caption{Optimization Configuration}
    \begin{tabular}{lcc}
    \toprule
    Hyper-parameter & Pre-training & Fine-Tuning \\
    \midrule
    Batch Size (\#patches/GPU) & 8k & 8k \\
    Peak Learning Rate & 1e-4 & 1e-5 \\
    Warm-up Steps & 10000 & 1000 \\
    Training Steps & 800k  & 50k \\
    Learning rate decay & Noam \citep{vaswani2017attention} & Noam \citep{vaswani2017attention} \\

    \bottomrule
    \end{tabular}
    \label{tab:optim_config}
\end{table}

\section{The Details of Experiments} \label{appendix:exp}
This section presents detailed experimental results on each task. In the following, if the training set and test sets come from different datasets, we label them as zero-shot settings. 
\subsection{TTS and VC tasks}
For TTS tasks, UniAudio is compared with the many previous SOTA models, Table \ref{tab:zero-shot-tts} presents the results. For FastSpeech 2, we only conduct QMOS evaluation as its implementation adopts speaker id as input \footnote{https://github.com/ming024/FastSpeech2}. 
We can see that UniAudio obtains better performance in terms of WER, SIM than YourTTS, VALL-E, NaturalSpeech 2 and Make-A-Voice. Compared with VoiceBox, UniAudio also gets comparable performance in terms of objective metrics. From the MOS evaluation, we can see that UniAudio can generate high-quality speech compared with previous SOTA works. Furthermore, UniAudio realizes the best zero-shot clone ability (\textit{e.g.} SMOS is 3.56 and SIM is 0.708). More experiments, such as cross-lingual zero-shot TTS and Mandarin Chinese speech synthesis can be found in demo page. For VC task, we conducted experiments on VCTK dataset, we randomly chose 200 audio pairs. PPG-VC and YourTTS are trained on small-scale datasets. Make-A-Voice and LM-VC \footnote{We seek help from the authors, they provide the inference results.} are trained on large-scale datasets as the same as UniAudio. Compared with previous work, UniAudio got better performance in voice conversion tasks.
\begin{table}[ht]
    \centering
    \small
    \caption{The performance comparison with previous SOTA methods in TTS and VC tasks. We do not conduct MOS evaluation for VALL-E, SPEARTTS and VoiceBox due to the models are not released.}
    \label{tab:zero-shot-tts}
    \scalebox{0.95}{
    \begin{tabular}{lccccc}
    \toprule
    Model     & Zero-shot & SIM ($\uparrow$) & WER ($\downarrow$) & MOS ($\uparrow$)   & SMOS ($\uparrow$)   \\
    \midrule
    \multicolumn{4}{l}{\textbf{Text-to-Speech}} \\
    GroundTruth      & -   &  - & 1.9  & 3.99$\pm$0.08   & - \\
    FastSpeech 2 \citep{fastspeech2}   & \xmark  & - & - &  3.81$\pm$0.10 & - \\
    YourTTS \citep{yourtts} & \cmark  & 0.337 & 7.7 &  3.66$\pm$0.07 & 3.02$\pm$0.07 \\
    VALL-E \citep{valle}  & \cmark   & 0.580 & 5.9  & - & - \\
    Make-A-Voice (TTS) \citep{makeavoice} & \cmark  &0.498 &5.7  & 3.74$\pm$0.08 & 3.11$\pm$0.06 \\
    NaturalSpeech 2 \citep{naturalspeech2} & \cmark   &0.620 & 2.3  & \textbf{3.83$\pm$0.10} & 3.11$\pm$0.10 \\
    SPEAR-TTS \citep{speartts} & \cmark   & 0.560 & / & - & - \\
    VoiceBox \citep{voicebox} & \cmark  &0.681 & \textbf{1.9}  & - & -  \\
    UniAudio  & \cmark  & \textbf{0.708} & 2.0  & 3.81$\pm$0.07 & \textbf{3.56$\pm$0.10} \\ 
    \midrule
    \multicolumn{4}{l}{\textbf{Voice Conversion}} \\
    GroundTruth   & -  &  -  & 3.25 & 3.74$\pm$0.08  & -   \\
    PPG-VC \citep{ppg-vc}  & \xmark & 0.78 & 12.3 & 3.41$\pm$0.10  & 3.47$\pm$0.10  \\
    YourTTS \citep{yourtts} & \cmark  & 0.719 & 10.1 &  3.61$\pm$0.10  & 3.26$\pm$0.10 \\
    Make-A-Voice (VC) \citep{makeavoice} & \cmark  & 0.678 & 6.2 & 3.43$\pm$0.09  & 3.47$\pm$0.10 \\
    LM-VC \citep{lm-vc} & \cmark  & 0.820  & 4.91 & 3.41$\pm$0.08  & 3.17$\pm$0.09  \\
    UniAudio  & \cmark  & \textbf{0.868} & {4.8} & 3.54$\pm$0.07  & \textbf{3.56$\pm$0.07} \\
    \bottomrule
    \end{tabular} 
    }
    \label{TTS}
\end{table}

\subsection{Speech Enhancement and Target Speaker Extraction}
For the SE task, we compare with previous SOTA methods, including discriminative methods (such as FullSubNet and FullSubNet+) and generative methods (such as SGMSE+ and NADiffuSE). Note that the CDiffuSE and NADiffuSE are both trained on the voicebank-demand dataset. Other models never saw the VCTK dataset in the training stage. We obtain the inference results based on their open-source models. Table \ref{tab:zero-shot-se} presents the results, we can see that UniAuido obtains the best DNSMOS score. The PESQ and VISQOL scores are lower than other SOTA methods, we think these metrics may not accurately assess the performance of generative methods. The similar finding is also observed in previous literature \citep{tokensplit} that the signal-level evaluation metrics may not be suitable for generative methods. In contrast, we recommend using DNSMOS and MOS scores as the main metrics. UniAuido can get good results in extremely noisy environments, we recommend readers refer to the demo page. For the TSE task, we conducted experiments on the LibriMix test set. The popular TSE systems: VoiceFilter \footnote{https://github.com/Edresson/VoiceSplit} and SpeakBeam\footnote{https://github.com/BUTSpeechFIT/speakerbeam} are used as baseline systems. As Table \ref{tab:zero-shot-se} shows, we can see that UniAudio obtains the best performance in terms of DNSMOS and MOS. 
\begin{table}[ht] 
    \centering
    \small
    \caption{The performance of SE and TSE tasks comparison with previous SOTA methods.}
    \label{tab:zero-shot-se}
     \scalebox{0.90}{
    \begin{tabular}{lcccccc}
    \toprule
    Model & Zero-shot  & PESQ ($\uparrow$) & VISQOL($\uparrow$) & DNSMOS($\uparrow$)   &MOS($\uparrow$) \\
    \midrule
    \multicolumn{4}{l}{\textbf{Speech Enhancement}} \\
    CDiffuSE \citep{lu2022conditional} & \xmark & 1.88  & 1.21  & 2.54   & - \\
    NADiffuSE \citep{wang2023nadiffuse} & \xmark & 2.96  & 2.41  & 3.03   & 3.30$\pm$0.08 \\
    SGMSE+ \citep{richter2023speech}  & \cmark & 3.21  & 2.72  & 3.29   & 3.56$\pm$0.08 \\
    FullSubNet \citep{hao2021fullsubnet}  & \cmark & 3.21  & 2.77  & 3.37  & 3.61$\pm$0.10 \\
    FullSubNet+ \citep{chen2022fullsubnet+} & \cmark  & 3.41  & 2.99  & 3.34   & 3.42$\pm$0.08 \\

    UniAudio  & \cmark  & 2.63  & 2.44  & \textbf{3.66} & \textbf{3.68$\pm$0.07} \\
    \midrule
    \multicolumn{4}{l}{\textbf{Target Speaker Extraction}} \\
    SpeakerBeam \citep{vzmolikova2019speakerbeam}  & \xmark    & 2.89  & 2.25  & 3.18   & 3.68$\pm$0.1 \\
    VoiceFilter \citep{wang2018voicefilter}  & \xmark     & 2.41  & 2.36  & 3.35  & 3.43$\pm$0.09 \\
    UniAudio & \cmark & 1.88  & 1.68  & \textbf{3.96}   & \textbf{3.72$\pm$0.06} \\
    \bottomrule 
    \end{tabular} }
    \label{vc}
    \vspace{-2mm}
\end{table}

\subsection{Singing Voice Synthesis}
Following Make-A-Voice, we conduct experiments on the M4Singer test set. We compare the generated singing samples with other systems, including 1) Diffsinger;  2) Make-A-Voice, a two-stage audio language model for singing voice generation. As illustrated in Table \ref{tab:zero-shot-sing}, we can see that UniAudio gets comparable results with Make-A-Voice and Diffsinger. 
\begin{table}[h] 
    \centering
    \small
    \caption{Quality and style similarity of generated samples in singing voice synthesis.}
    \label{tab:zero-shot-sing}
    \begin{tabular}{lcc}
    \toprule
    Model  & MOS ($\uparrow$)     & SMOS ($\uparrow$)       \\
    \midrule
    Diffsinger \citep{liu2022diffsinger}    &  3.94$\pm$0.02 & \textbf{4.05$\pm$0.06}   \\
    Make-A-Voice \citep{makeavoice}   &  3.96$\pm$0.03 & 4.04$\pm$0.05    \\
    UniAudio & \textbf{4.08$\pm$0.04}   & 4.04$\pm$0.05   \\
    \bottomrule 
    \end{tabular} 
    \label{svs}
\end{table}
\subsection{Text-to-sound and text-to-music generation}
The text-to-sound generation task has attracted great interest in audio research. Following Diffsound \citep{diffsound}, most of the methods evaluate their systems on the AudioCaps \citep{audiocaps} test set. However, we found that if the training data includes the AudioCaps data, the model is easy to overfit with AudioCaps. As a result, the best performance can be obtained when the model only trains on the Audiocaps. In this study, we conduct a zero-shot evaluation on the Cloth test set \citep{drossos2020clotho}. Table \ref{tab:zero-shot-sound} shows the results. We can see that UniAudio obtains better performance than Diffsound and AudioLDM. Compared to recent SOTA models, such as Tango and Make-an-Audio 2, UniAudio also gets comparable performance. For the text-to-music task, we follow MusicGen \citep{musicgen}, evaluating our methods on MusicCaps \citep{musiclm}. Compared with previous SOTAs, UniAudio gets a comparable performance with other models. From the MOS evaluation performance, we can see that MusicGen is better than our current models. We speculate one of the reasons is that MusicGen uses a large-scale high-quality dataset (20k hours). 
\begin{table}[h] 
    \centering
    \small
    \caption{Text-to-sound and text-to-music evaluation. We report the subjective metrics including FAD($\downarrow$), and KL($\downarrow$). Furthermore, we also conduct objective evaluation. Note that the training data of AudioGen includes Cloth datatset, thus can not be seen as zero-shot setting.}
    \label{tab:zero-shot-sound}
    \begin{tabular}{cc|cc|cc}
    \toprule    
    Model &Training Data (Hours) & FAD  & KL   & OVL.  & REL.  \\
    \midrule
    \multicolumn{4}{l}{\textbf{Text-to-Sound Generation}} \\
    Reference  & /  &  / &/    & 70.47$\pm$1.9  & 78.84$\pm$1.5\\
    Diffsound & 2k  & 7.8   &  6.53  & -  & -   \\
    AudioGen & 4k & 2.55   &  2.5  &63.84$\pm$2.1   & \textbf{72.12$\pm$1.8}   \\
    Tango  & 3.3k    & 3.61   &  2.59  & \textbf{66.2$\pm$1.7}   & 68.57$\pm$1.5   \\
    Make-an-Audio 2  & 8.7k  &2.13    & 2.49   & 61.52$\pm$1.6 &  69.9$\pm$1.5  \\
    AudioLMD & 9k  & 4.93   &  2.6  &60.95$\pm$1.9   & 65.7$\pm$1.8  \\
    UniAudio &7k & 3.12   &  2.57                  
        &61.9$\pm$1.9   & 66.1$\pm$1.5 \\
    \midrule
    \multicolumn{4}{l}{\textbf{Text-to-Music Generation}} \\
    Riffusion  & -  & 14.8 & 2.06 & - & - \\
    Mousai  & - & 7.5   & 1.59   &-   & - \\
    MusicLM & 280k  & 4.0  & -   &-   & - \\
    Noise2Music & 280k  & 2.1  & -   & -   & - \\
    MusicGen  & 20k & 4.52  & 1.41   & \textbf{73.28$\pm$1.5}   & \textbf{71.28$\pm$1.7} \\
    UniAudio & 8k & 3.65   &  1.87             
        &67.85$\pm$1.70   &  70.0$\pm$1.5 \\
    \bottomrule
    \end{tabular}
    \label{table:text-to-audio}
\end{table}

\subsection{Audio Edit}
Audio edit aims to edit the original audio based on Human's instruction. 
AUDIT \citep{wang2023audit} is the SOTA model in audio edit task, which designs a data simulation strategy to get triplet training and test data (\textit{e.g.}, \{audio, audio, text\}). The authors set 5 different tasks, including adding, dropping, replacing, inpainting and super-resolution, and simulated large-scale data for each task. To validate that our pre-trained model can be fine-tuned with small-scale data, we choose adding, dropping and super-resolution tasks to fine-tune simultaneously. To finish the fine-tuning process, we define a new task label: \textit{Audit\_task}. The experimental results as Table \ref{table:audit} shows. We can observe that: (1) UniAudio can get better performance with the previous SOTA model. (2) Fine-tuning pre-trained UniAudio can get better performance than training it from scratch, which further validates the effectiveness of pre-training a model on large-scale training data.

\begin{table}[h]
    \centering
    \small
    \caption{Audio edit task evaluation.}
     \label{table:audit}
    \begin{tabular}{cccc}
    \toprule    
    Type & Model  & FD  & KL   \\
    \midrule
    \multicolumn{2}{l}{\textbf{Adding task}} \\
    & AUDIT  & 21.80 &  \textbf{0.92}   \\
    \midrule
    & UniAudio (scratch)  & 20.2 &  0.99  \\
    & UniAudio (fine-tune)  & \textbf{19.69} &  {0.934}  \\
    \midrule
    \multicolumn{2}{l}{\textbf{Dropping task}} \\
    & AUDIT  & 22.40 &  0.95   \\
    \midrule
    & UniAudio (scratch)  & 27.76 &  1.38  \\
    & UniAudio (fine-tune)  & \textbf{23.1} &  \textbf{1.10}  \\
    \midrule
    \multicolumn{2}{l}{\textbf{Super-Resolution task}} \\
    & AUDIT  & 18.14 &   0.73   \\
    \midrule
    & UniAudio (scratch)  & 11.51 &  0.29  \\
    & UniAudio (fine-tune)  & \textbf{10.54} &  \textbf{0.289}  \\
    \bottomrule
    \end{tabular}
\end{table}

\begin{table}[h]
    \centering
    \small
    \caption{Quality and style similarity of generated samples for Instructed TTS task.}
    \label{tab:prompt-tts}
    \begin{tabular}{lcc}
    \toprule
    Model   & MOS ($\uparrow$)     & SMOS ($\uparrow$)        \\
    \midrule
    GT        & 3.77$\pm$0.07  &  3.85$\pm$0.08 \\
    UniAudio (scratch) &  3.62$\pm$0.07 & 3.67$\pm$0.08   \\
    UniAudio (tuning)&  3.61$\pm$0.09 & 3.71$\pm$0.09 \\
    \bottomrule 
    \end{tabular}
    \label{svs}
\end{table}

\subsection{Instructed TTS}
Using instruction to guide speech synthesis has received great attention \citep{guo2023prompttts,yang2023instructtts}. In this part, we fine-tune the UniAudio model on the PromptSpeech \citep{guo2023prompttts} dataset. Furthermore, we also try to train a UniAudio model from scratch with the PromptSpeech dataset. Different from previous works that designed special style encoders to capture the style information from text descriptions, we directly use the T5 text encoder to extract representations from text and then combine it with the phoneme sequence input to the UniAudio, which is more convenient.\footnote{Note that the authors of PromptTTS \citep{guo2023prompttts} told us their objective metrics tools, checkpoints, and generated samples have been lost due to the machine errors. Thus we cannot fairly compare with them.} Table \ref{tab:prompt-tts} shows the results, we can see that UniAudio has good performance in terms of style control and speech quality when compared with the ground truth samples.

\subsection{Speech Dereverberation}
For the speech dereverberation task, we use the Room Impulse Response (RIR) data from the openSLR26 and openSLR28 dataset, and the speech data from the LibriTTS clean part. We simulate about 100 hours of training data and 1 hour of test data. We compare with previous SOTA systems, such as FullSubNet, FullSubNet+ and SGMSE+. Table \ref{tab:zero-shot-sd} presents the results. We can see that UniAudio obtains the SOTA performance in speech dereverberation tasks with small-scale training data in terms of DNSMOS metric. Similar with speech enhancement task, we speculate that PESQ may not suitable for the generative methods.  
\begin{table}[h]
    \centering
    \small
    \caption{Results comparison with previous speech Dereverberation systems.}
    \label{tab:zero-shot-sd}
    \begin{tabular}{lcc}
    \toprule
    Model   & PESQ ($\uparrow$) & DNSMOS($\uparrow$)  \\
    \midrule
    SGMSE+   & 2.87   & 3.42   \\
    FullSubNet   & 2.29    & 3.32  \\
    FullSubNet+   & 2.27    & 3.25   \\
    UniAudio (scratch)   & 1.23    & 3.18   \\
    UniAudio (tuning)   & 2.13    & \textbf{3.51}   \\
    \bottomrule 
    \end{tabular}
    \label{vc}
    \vspace{-2mm}
\end{table}

\subsection{Speech Edit}
For the speech edit task, we use the LibriTTS dataset. In practice, we randomly choose some words to mask in the training stage. We expect the model to recover the whole speech based on the phoneme sequence. In the inference stage, we can mask the region that we want to update in the speech and input the new words so that the model can edit the speech. For this task, we take the TTS system that regenerates a complete waveform from the whole sentence to be edited as the baseline. In the evaluation, we mainly validate three situations: (1) word replacement; (2) insert a new word; and (3) delete a word. For each situation, we randomly chose 10 sentences from the LibriTTS test clean set.

\section{Ablation study} \label{ab: ablation study}

\subsection{The influence of multi-task training} \label{ablation: multi-task}
In this part, we explore whether multi-task training can bring better performance than task-specific training. To answer this question, we use the same model trained on different tasks, respectively. Table \ref{tab:ablation-multi-task} shows the experimental results, UniAudio (single) means that the model is trained on a single task. We observe that multi-task training brings the gain over all of the tasks. In Appendix \ref{why work}, we give some potential reasons why multi-task training can bring improvement. 
\begin{table}[h]
    \centering
    \caption{The ablation study of the effectiveness of multi-task training.}
    \scalebox{0.68}{
    \begin{tabular}{cccccc}
    \toprule
    \multirow{2}{*}{Task} & \multirow{2}{*}{Model} &  \multicolumn{2}{c}{Objective Evaluation}  & \multicolumn{2}{c}{Subjective Evaluation} \\  
    && Metrics & Results & Metrics & Results \\
    \midrule
    \multirowcell{2}{Text-to-Speech} & UniAudio (Single) & \multirow{2}{*}{SIM$(\uparrow)$ / WER$(\downarrow)$} & 0.64 / 2.4 & \multirowcell{2}{MOS$(\uparrow)$ \\/ SMOS$(\uparrow$)} & {3.77$\pm$0.06} / 3.46$\pm$0.10 \\ 
    & UniAudio & & \textbf{0.71 / 2.0} & & \textbf{
    3.81$\pm$0.07} / \textbf{3.56$\pm$0.10} \\
    \hline
    \multirowcell{2}{Voice \\ Conversion} & UniAudio (Single) & \multirowcell{2}{SIM$(\uparrow)$ / WER$(\downarrow)$} & 0.84 / 5.4 & \multirowcell{2}{MOS$(\uparrow)$ \\/ SMOS$(\uparrow$)} & 3.45$\pm$0.07 / 3.44$\pm$0.07 \\ 
    & UniAudio & & \textbf{0.87 / 4.8} & & \textbf{3.54$\pm$0.07 / 3.56$\pm$0.07} \\
    \hline
    \multirowcell{2}{Speech \\ Enhancement} & UniAudio (Single) & \multirowcell{2}{PESQ$(\uparrow)$ \\ / VISQOL$(\uparrow)$ / DNSMOS$(\uparrow)$}& 2.35 / 2.30 / {3.45} & \multirow{2}{*}{ MOS$(\uparrow$)} & {3.65$\pm$0.08} \\ 
    & UniAudio &  & \textbf{2.63} / \textbf{2.44} / \textbf{3.66} & & \textbf{3.68$\pm$0.07} \\
    \hline
    \multirowcell{2}{Target Speaker \\ Extraction} & UniAudio (Single) & \multirowcell{2}{PESQ$(\uparrow)$ \\ / VISQOL$(\uparrow)$ / DNSMOS$(\uparrow)$} & \textbf{1.97} / 1.61 / {3.93} & \multirow{2}{*}{ MOS$(\uparrow$)} & {3.58$\pm$0.08} \\ 
    & UniAudio &  & 1.88 / \textbf{1.68} / \textbf{3.96} & & \textbf{3.72$\pm$0.06} \\
    \hline
    \multirowcell{2}{Singing Voice \\ Synthesis} &  UniAudio (Single) & \multirow{2}{*}{-} & \multirow{2}{*}{-} & \multirowcell{2}{MOS$(\uparrow$) \\/ SMOS$(\uparrow$)} & \textbf{4.14$\pm$0.07} / 4.02$\pm$0.02 \\ 
    & UniAudio &  &  & & {4.08$\pm$0.04} / \textbf{4.04$\pm$0.05} \\
    \hline
    \multirowcell{2}{Text-to-Sound} & UniAudio (Single) & \multirow{2}{*}{FAD $(\downarrow)$ / KL $(\downarrow)$} & {3.84 / 2.7} & \multirowcell{2}{OVL $(\uparrow)$ \\/ REL $(\uparrow)$} & {60.0$\pm$2.1 / 61.2$\pm$1.8} \\ 
    & UniAudio &  & \textbf{3.12 / 2.6} & & \textbf{61.9$\pm$1.9 / 66.1$\pm$1.5} \\
    \hline
    \multirowcell{2}{Text-to-Music} & UniAudio (Single) & \multirow{2}{*}{FAD $(\downarrow)$ / KL $(\downarrow)$} & 5.24 / \textbf{1.8} & \multirowcell{2}{OVL $(\uparrow)$ \\/ REL $(\uparrow)$} &  64.4$\pm$2.1 / 66.2$\pm$2.4 \\
    & UniAudio &  & \textbf{3.65} / 1.9 & & \textbf{67.9$\pm$1.7 / 70.0$\pm$1.5} \\
    \hline
    \multirowcell{2}{Audio Edit} & UniAudio (single) & \multirow{2}{*}{FD $(\downarrow)$ / KL $(\downarrow)$} & 19.82 / {0.92} & \multirowcell{2}{-} &  - \\
    & UniAudio &  & \textbf{17.78 / 0.77} & & - \\
    \hline
    \multirowcell{2}{Speech Dereverb.} & UniAudio (single) & \multirow{2}{*}{PESQ$(\uparrow)$ / DNSMOS$(\uparrow)$} & 1.23 / 3.18 & \multirowcell{2}{-} &  - \\
    & UniAudio &  & \textbf{2.13 / 3.51} & & - \\
    \hline
    \multirowcell{2}{Instructed TTS} & UniAudio (single) & \multirow{2}{*}{-} & - & \multirowcell{2}{MOS$(\uparrow)$ / SMOS$(\uparrow)$} &  \textbf{3.62$\pm$0.07} / 3.67$\pm$0.08 \\
    & UniAudio &  & - & & 3.61$\pm$0.09 / \textbf{3.71$\pm$0.09}  \\
    \hline
    \multirowcell{2}{Speech Edit} & UniAudio (single) & \multirow{2}{*}{MCD $(\downarrow)$} &  5.26 & \multirowcell{2}{MOS$(\uparrow)$} &  {3.73$\pm$0.07} \\
    & UniAudio &  & \textbf{5.12} & & \textbf{3.82$\pm$0.06} \\
    \bottomrule 
    \vspace{-10pt}
    \end{tabular}}
    \label{tab:ablation-multi-task}
\end{table}
\subsection{Fine-tuning the pre-trained model on the new task will influence the performance on previous tasks?} \label{ablation: fine-tune}
\begin{figure*}[h]
    \centering
    \includegraphics[width=\textwidth]{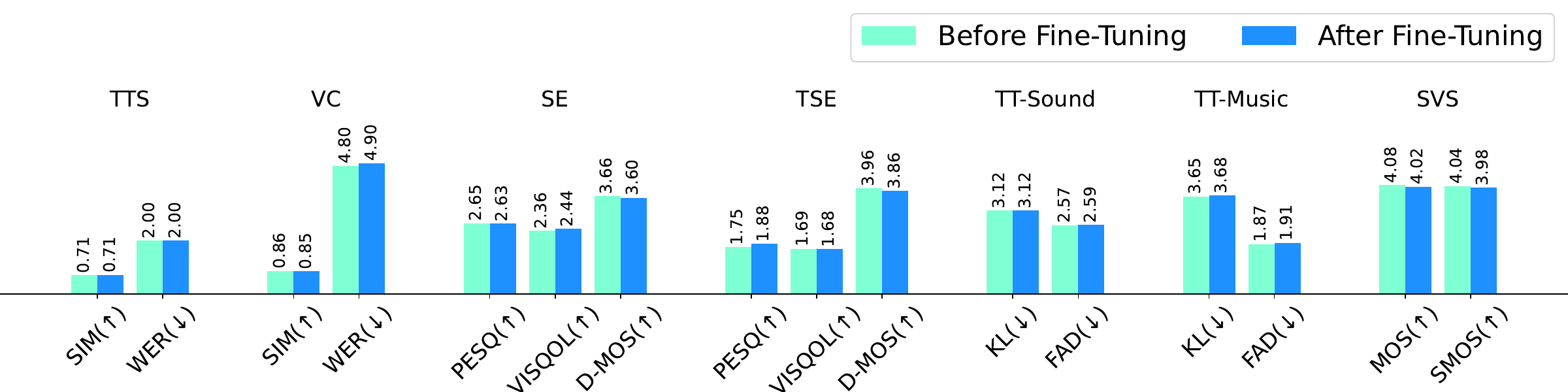}
    \caption{Performance comparison over 7 audio generation tasks before/after fine-tuning.}
    \label{fig:finue}
\end{figure*}

In this part, we conduct experiments to explore whether fine-tuning the pre-trained model on new tasks will influence the performance of previous tasks. We evaluate the pre-trained UniAudio model (trained on 7 tasks) and fine-tuned UniAudio model (fine-tuned on 4 new tasks) on 7 tasks. Figure \ref{fig:finue} shows the results. We can see that the performance does not significantly drop on previous training tasks, which demonstrates that UniAudio has the potential to add new tasks continuously without losing previous task knowledge.

\subsection{The influence of data quantity} \label{ablation: data-quantify}
\begin{figure*}[t]
    \centering
    \includegraphics[width=\textwidth]{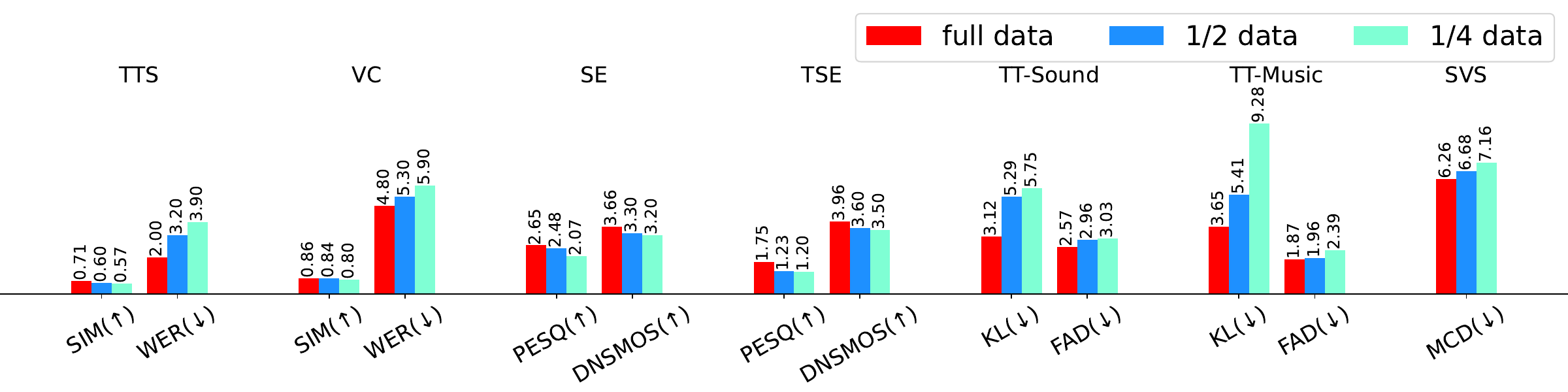}
    \caption{Performance comparison over different data quantity.}
    \label{fig:data-quantify}
\end{figure*}

In this part, we conduct experiments to explore the influence of data quantity, we give three settings: (1) using all of the data; (2) using $1 / 2$ training data for each task; (3) using $1 / 4$ training data for each task. We present the results in Figure \ref{fig:data-quantify}. Based on the experimental results, this work claims that the data quantity is a key point to building a strong audio foundation model. In the future, we will explore to use of more unlabeled data to help improve the performance.

\section{Why UniAudio Can Work Well?} \label{why work}
From the previous discussions, we can see that the universal modeling strategy brings improvement for different tasks. In this part, we try to give some potential explanations. 

(1) \textbf{Deterministic latent space}: we formulate different modalities into a deterministic latent space (fixed vocabulary) by tokenization. Different tokens can be seen as specific 'words', and we can use a next-token prediction strategy to train the model. Similar to GPT-series \citep{radford2018improving,radford2019language}, such strategy creates the opportunity for the model to learn the intrinsic properties of audio and the interrelationship between audio and other modalities. 

(2) \textbf{Shared information between different types of audio}: Although multiple types of audio (speech, sounds, music, and singing) present significant differences in the time domain or frequency domain, neural audio codec models effectively capture their shared information (rethinking the working principle of neural codecs, which similar information will be allocated the same token id). Due to the shared information that exists in different types of audio, multi-task training can be seen as increasing training data for each task. 

(3) \textbf{Data augmentation perspective}: We speculate that multi-task training can be viewed as data augmentation for some tasks. Considering the TTS and VC task's definition: \\
\texttt{TTS:  <phoneme\_sequence> <prompt> <audio\_sequence>} \\
\texttt{VC:  <semantic\_token> <prompt> <audio\_sequence>} \\
We can see that the difference in task formulation for TTS and VC is that they use different ways to denote the phonetic information. In essence, they carry the same phonetic information. The difference is that semantic tokens include the duration information. Thus we can view the phoneme sequence as a special semantic sequence that drops the duration information. Such dropping operation is widely used as a data augmentation strategy \citep{specaugment}. 



\section{The details of Audio Codec Models} \label{appendix: codec}
\begin{table}[h]
    \centering
    \small
    \vspace{-5mm}
    \caption{Performance comparison between encodec and our universal neural codec. FPS: frame per second; TPS: token per second. Perceptual evaluation of speech quality (PESQ$\uparrow)$; Short Term Objective Intelligibility (STOI$\uparrow$).}
    \vspace{2mm}
    \scalebox{0.7}{
    \begin{tabular}{lccc|cc|cc|cc|cc|cc}
    \toprule
    \multirow{2}{*}{Type} & &&&  \multicolumn{2}{c}{Speech (VCTK)} & \multicolumn{2}{|c}{Sound (cloth)}  & \multicolumn{2}{|c}{Music (musiccaps)} & \multicolumn{2}{|c}{Sing (m4sing)}  & \multicolumn{2}{|c}{Average} \\ 
    &&&& \multicolumn{2}{c|}{\citep{vctk}} & \multicolumn{2}{c|}{\citep{drossos2020clotho}} & \multicolumn{2}{c|}{\citep{musiclm}} & \multicolumn{2}{c|}{\citep{m4singer}} & \multicolumn{2}{c}{-}\\
    Model   & $n_q$ & FPS & TPS & PESQ  & STOI   & PESQ  & STOI   & PESQ  & STOI   & PESQ  & STOI   & PESQ  & STOI  \\
    \midrule
    Encodec  & 8 & 75 & 600 & 2.18  & 0.79 & 2.23  & 0.48 & 1.86  & {0.57} & 1.95  & 0.76 & 2.05  & 0.65        \\
    Ours   & {3} & {50} & 150 & 2.96 & 0.85 & {2.42} & {0.49} & {1.99} & {0.57} & {3.13} & {0.85} & {2.62} & {0.69}     \\
    Ours   & 4 &50 & 200  & 3.11 & 0.86 & 2.5 & 0.51 & 2.08 & 0.59 & 3.27 & 0.86 & 2.73 & 0.71       \\
    Ours   & 8 &50 &400  & \textbf{3.36} &  \textbf{0.88} & \textbf{2.67} &  \textbf{0.54} & \textbf{2.31} &  \textbf{0.65} & \textbf{3.49} &  \textbf{0.89} & \textbf{2.95} & \textbf{0.74}       \\
    \bottomrule 
    \end{tabular}}
    \label{tab:all-codec-comparison}
\end{table}

\begin{figure*}[h]
    \centering
    \includegraphics[width=0.8\textwidth]{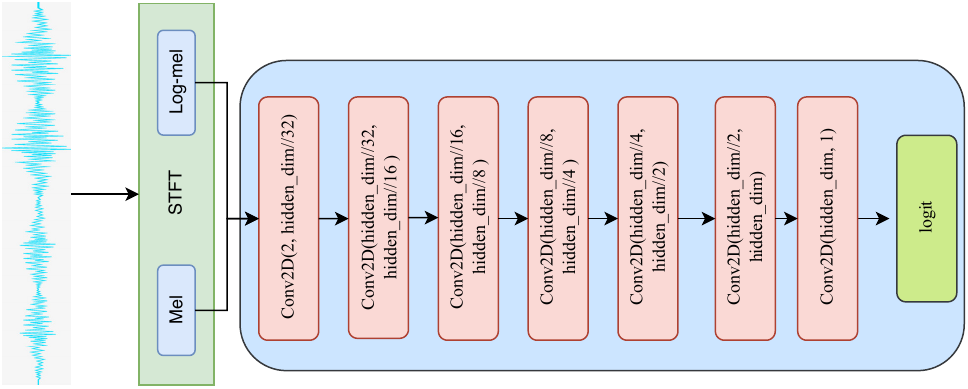}
    \caption{The overview of a single Mel-based discriminator. In practice, we will use multiple discriminators by setting different hop lengths and hidden dimensions.}
    \label{fig:codec}
  \end{figure*}

In this part, we give more details about our neural audio codec model in Section \ref{sec:audio}. We adopt a similar encoder-decoder framework with the Encodec model, the difference includes: (1) we replace the multi-scale STFT-based (MS-STFT) discriminator as our multi-scale Mel-based discriminator. (2) We rewrite the vector quantization implementation \footnote{Please refer to our source code to find the details.} based on Encodec's open-source version \footnote{https://github.com/facebookresearch/encodec/blob/main/encodec/quantization/core\_vq.py}, making it more suitable for DDP training. Figure \ref{fig:codec} shows the details of the mel-based discriminator. We combine the mel-spectrogram and log-mel-spectrogram features and then input them into a network consisting of several convolutional layers. Our motivation is that the mel-spectrogram has a strong intrinsic inductive bias, especially for sounds and music-related audio (the SOTA sounds or music classification systems are based on the log-mel-spectrogram in the literature.). Thus, we speculate that choosing a mel-spectrogram-based discriminator can better promote high-fidelity audio reconstruction. In our experiments, we use 6 different discriminators with different configurations \footnote{In our experiments, we find the mel-based discriminator brings better reconstruction performance when we train a universal neural audio codec.}. Specifically, we set the hidden\_dim as \{64, 128, 256, 512, 512, 512\} and the hop length as \{32, 64, 128, 256, 512, 1024\}. We train the neural audio codec model based on the Librilight and AudioSet datasets. 
Table \ref{tab:all-codec-comparison} demonstrates that the neural codec model adopted in this work outperforms prior Encodec \citep{encodec}.

\section{Subjective Evaluation} \label{appendix: evaluation metric}
For TTS and VC tasks, we focus on speech quality (QMOS) and speaker similarity (SMOS). The details are as follows.
For speech quality evaluation, we conduct the MOS (mean opinion score) tests and explicitly ask the raters to \textit{focus on examining the audio quality and naturalness, and ignore the differences of style (timbre, emotion, and prosody}. The testers present and rate the samples, and each tester is asked to evaluate the subjective naturalness on a 1-5 Likert scale.

For speaker similarity evaluation, we ask the raters to \textit{focus on the similarity of the speaker identity (timbre) to the reference, and ignore the differences in content, grammar, or audio quality}. We paired each synthesized utterance with a reference utterance to evaluate how well the synthesized speech matched that of the target speaker.

\begin{figure*}[t]
    \centering
    \includegraphics[width=1.0\textwidth]{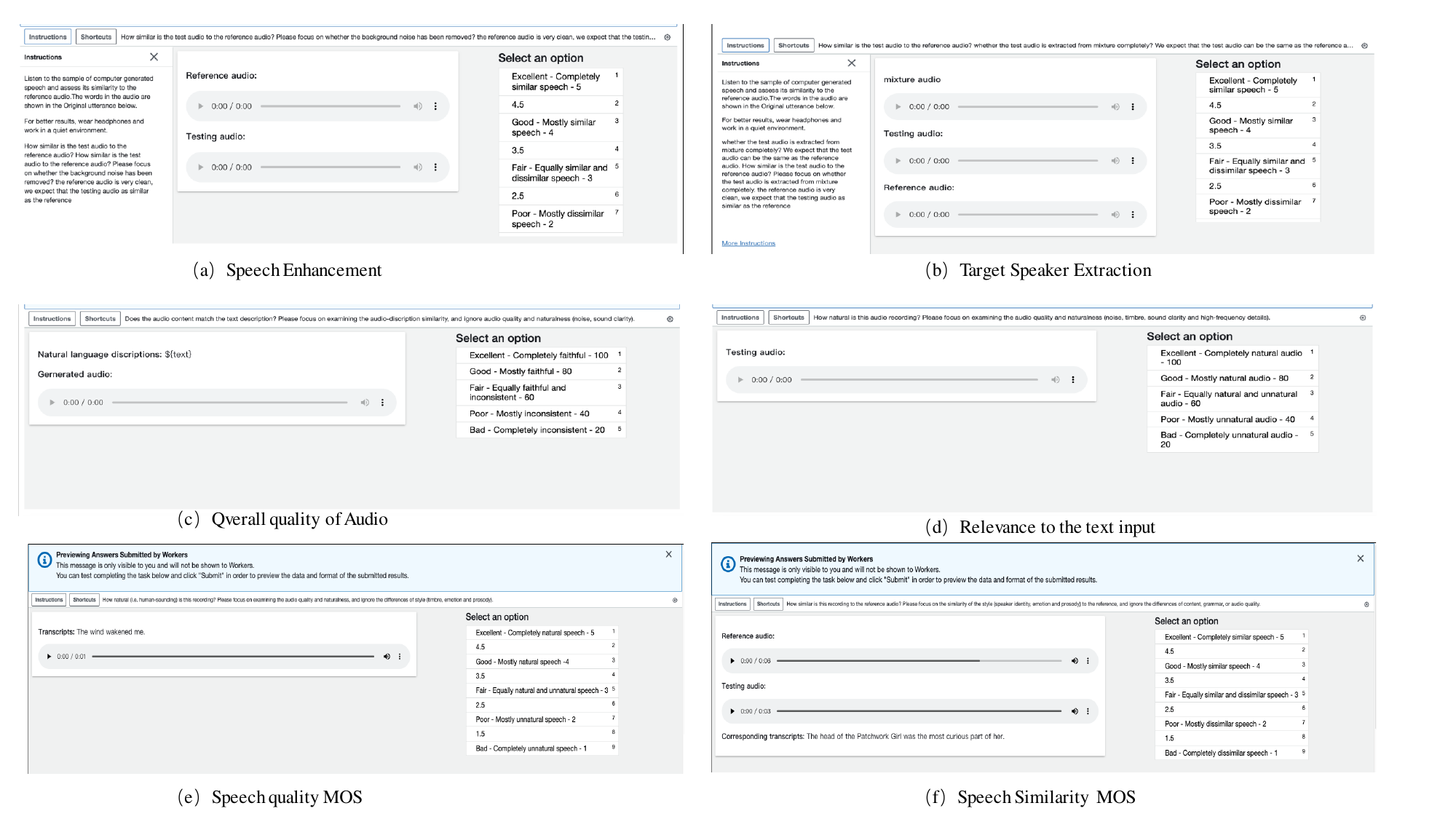}
    \caption{Screenshots of subjective evaluations.}
    \label{fig:mos}
  \end{figure*}

For SE and TSE tasks, we write explicit instructions to ask the rater to assess the generated speech. Refer to Figure \ref{fig:mos} to see the details.

For SVS, we also conduct quality MOS (QMOS) and style similarity MOS (SMOS). Different from TTS's SMOS evaluation,  we explicitly instruct the raters to \textit{focus on the similarity of the style (timbre, emotion, and prosody) to the reference, and ignore the differences in content, grammar, or audio quality}.  

For sound and music generation tasks, we follow AudioGen \citep{kreuk2022audiogen} and MusicGen \citep{musicgen} to evaluate (1) overall quality (OVL), and (2) relevance to the text input (REL).

Our subjective evaluation tests are crowd-sourced and conducted by 20 native speakers via Amazon Mechanical Turk. The screenshots of instructions for testers have been shown in Figure~\ref{fig:mos}. We paid about \$500 on participant compensation. A small subset of speech samples used in the test is available at \url{https://uniaudio666.github.io/demo_UniAudio/}.

\end{document}